# Local probing of superconductivity at oxide interfaces with atomic force microscopy

Dilek Yildiz[1,2,3,*,†], Sungmin Kim[1,2,*], Dengyu Yang[1,2,4,*], Muqing Yu[4,*], Kyoungjun Lee[5], Ruiqi Sun[5], En-Min Shih[1,6], Steven R. Blankenship[1], Patrick Irvin[4], Franz J. Giessibl[7], Chang-Beom Eom[5], Jeremy Levy[4], and Joseph A. Stroscio[1,†]

[1]Physical Measurement Laboratory, National Institute of Standards and Technology; Gaithersburg, MD 20899, USA.
[2]Joint Quantum Institute, Department of Physics, University of Maryland; College Park, MD 20742, USA
[3]Department of Advanced Material Science, The University of Tokyo; Chiba 277-8561, Japan.
[4]Department of Physics and Astronomy, University of Pittsburgh; Pittsburgh, Pennsylvania 15260, USA
[5]Department of Materials Science and Engineering, University of Wisconsin–Madison; Madison, Wisconsin 53706, USA
[6]Department of Chemistry and Biochemistry, University of Maryland; College Park, MD 20742, USA.
[7]Institute of Experimental and Applied Physics, University of Regensburg; Regensburg 93040, Germany

[*]These authors contributed equally to this work.

[†]Corresponding authors: yildiz@g.ecc.u-tokyo.ac.jp, joseph.stroscio@nist.gov

**Abstract:** Superconductivity in strontium titanate has remained enigmatic for over 50 years. The LaAlO$_3$/SrTiO$_3$ (LAO/STO) heterointerface enables systematic dimensional confinement, from two-dimensional electron gas to quasi-one-dimensional nanostructures, providing unprecedented access to this quantum state. While transport measurements in patterned devices reveal puzzling phenomena, including width-independent critical currents and anomalous pairing, suggesting one-dimensional behavior, direct local probes for the patterned interface and its superconducting response have remained lacking. Here, we use ultra-low temperature non-contact atomic force microscopy with dissipation spectroscopy and Kelvin probe force microscopy to locally probe signatures of superconductivity in patterned LAO/STO devices. Spatially resolved energy dissipation measurements reveal signatures of superconductivity with some features confined to edge channels of order ≈200 nm in width. Dissipation spectra exhibit a characteristic nonlinear bias dependence that provides a local diagnostic of superconductivity consistent with the intermediate carrier density near the superconducting dome, which persists up to the critical field. These results demonstrate the ability of atomic force microscopy to probe superconductivity in patterned LAO/STO structures, potentially addressing fundamental, longstanding questions about quantum confinement and transport anomalies in these correlated nanostructures.

**Main Text:**

One of the significant technological advancements of this century is the development of semiconductor nanoelectronics, which has opened new possibilities for quantum information technologies. Beyond fabricating nanoscale electronic devices, these platforms are essential for advancing our understanding of complex quantum systems for investigating key physical phenomena, such as superconductivity and strong electronic correlations. STO, the first known superconducting semiconductor, provides a unique platform for such studies, offering a convenient route to integrate superconductivity into mesoscopic devices. It has a bandgap of 3.2 eV, but becomes superconducting below 300 mK[1]. The LAO/STO heterointerface[2] has revolutionized our capability to probe these mysteries, creating a tunable two-dimensional electron system (2DES) that inherits superconductivity from bulk STO[3]. Through conductive atomic force microscope (c-AFM) lithography[3,4] and ultra-low-voltage electron beam lithography (ULV-EBL)[5,6], this 2DES can be programmed with sub-10 nm resolution, creating quasi-one-dimensional structures that exhibit remarkable quantum phenomena.

A central unresolved question is whether superconductivity in confined STO geometries retains its two-dimensional nature or becomes fundamentally one-dimensional, a distinction with profound implications for both fundamental physics and quantum device engineering. Transport measurements in these nanostructured devices have revealed a constellation of anomalous behaviors that defy conventional understanding of superconductivity. Most strikingly, the critical current remains independent of channel width[7], a violation of the basic scaling laws expected for two-dimensional superconductors. Electron pairing persists far outside of the superconducting regime[8], and devices exhibit a Pascal conductance sequence[9] indicating strong pairing of $n = 2, 3, 4,…$ electrons within 1D channels. Collectively, these observations suggest that the pairing glue and associated superconductivity in patterned LAO/STO is fundamentally one-dimensional, confined to edge channels regardless of the nominal device geometry.

To date, direct probing of the superconductivity at the interface has remained challenging at ultra-low temperatures. Moreover, microscopic evidence of one-dimensional superconducting behavior has remained elusive. Despite extensive studies suggesting such behavior, these transport measurements average over device dimensions and cannot resolve the spatial structure of the superconducting state. Previous scanning probe efforts, including scanning quantum interference device (SQUID)[10], scanning single-electron transistor (SET)[11], near-field optical microscopy[12], and microwave impedance microscopy[13], have revealed important aspects of the LAO/STO system, but none have directly imaged superconductivity at the nanoscale. Crucially, none of these techniques was performed at the ultra-low temperatures necessary to probe the superconducting state ($T_c$ < 300 mK), nor did they possess the spatial resolution and energy sensitivity required to distinguish one-dimensional from two-dimensional superconducting channels. This knowledge gap has left fundamental questions unanswered: Where precisely is superconductivity localized within patterned structures? Are the channels truly one-dimensional? How does spatial confinement relate to the anomalous transport properties?

Here, we directly probe the superconducting state in patterned LAO/STO nanostructures using ultra-low temperature (10 mK) non-contact atomic force microscopy (AFM) and Kelvin probe force microscopy (KPFM) with atomically sharp tips. This approach enables us to image superconducting channels with nanometer resolution while simultaneously measuring local energy dissipation—a sensitive probe of the superconducting state. Non-contact dissipation spectroscopy is uniquely suited to detect the suppression of ohmic losses characteristic of superconductivity and the changes in the dominant dissipation mechanisms channels between normal and superconducting regions[14,15,16,17,18]. By correlating spatial maps of contact potential difference with magnetic-field-dependent dissipation measurements, we directly visualize where superconductivity resides within nanometer-scale patterned structures at the oxide interface.

**Device patterning and macroscopic transport characterization**

Using established nanoscale patterning methods (see Supplementary Information), c-AFM and ULV-EBL are used to pattern LAO/STO interfaces into superconducting devices, including nanowires and electron waveguides, for investigation with ultra-low-temperature scanning probe techniques. Device A with 4 unit cells (uc) of LAO is patterned by ULV-EBL (Fig. 1a), while Device B (6 uc LAO) is patterned by c-AFM (Supplementary Fig. 2a). The patterned regions are highlighted in green color in Fig. 1c-d. Each patterned device contains three distinct channel geometries: a two-dimensional sheet, a one-dimensional wire, and a waveguide channel (Fig. 1d). Device conductance is monitored while patterning; a sudden increase in conductance is observed when a nanowire is written between two electrodes, demonstrating successful selective switching of the interface from insulating to conducting (Fig. 1b and Supplementary Fig. 2b).

After patterning, devices are then carefully transported in vacuum to the millikelvin SPM system in order to minimize degradation of the conducting patterns due to atmospheric exposure (see Supplemental Information). After cooling the device to cryogenic temperatures (10 mK), we measured its transport properties to benchmark the quality of the nanoscale patterning. The current-voltage (*I-V*) curves directly measure superconducting properties, and the differential resistance (*dV/dI*) as a function of current and magnetic field is examined to extract the critical current and field. Figure 1e and f show the corresponding measurements for (Device A, 4 uc, ULV-EBL) patterned channels, with the current applied between the upper electrodes 4 and 5 at 10 mK (see Fig. 1d). Similar measurements were made for a second device patterned by c-AFM (Device B, Supplementary Fig. 4). Both devices show a superconducting critical current ($I_C$) of 30 nA. Critical fields are also determined in both devices, corresponding to the suppression of superconductivity.

A four-terminal transconductance map is recorded by varying the source-drain bias and the side-gate voltage, revealing the finite-bias spectroscopy measurements from zero to finite magnetic fields (Supplementary Fig. 4c). These results are consistent with similar nanostructures fabricated in previous measurements, indicative of the quality of the patterned 2DES[19,20]. In summary, the transport measurements confirm the superconducting nature of the patterned channels in both devices, consistent with previous reports.

**Scanning probe microscopy measurements**

Local measurements were performed with frequency-modulation non-contact AFM, providing spatially resolved topographic and electrostatic information (see Supplemental Information). Kelvin probe force microscopy (KPFM) measures the local potential landscape of the LAO surface by determining the contact potential difference (CPD) between the LAO/STO surface and the AFM probe, as schematically shown in Fig. 2a and b. We first investigated the patterned regions in the ULV-EBL-patterned device (Device A). In the non-contact AFM topography scans, the patterned structures are clearly visible with strong contrast (Fig. 2c). We attribute the observed contrast to differences in tip–sample interactions between the patterned and unpatterned regions, arising from variations in surface work function and local carrier concentration within the 2DES. Specifically, the change in CPD will change the frequency shift ($\Delta f$) vs bias voltage curves (Supplementary Fig. 1). Therefore, at fixed bias, variations in $\Delta f$ result in corresponding changes in the measured height (*Z*) to keep $\Delta f$ constant in the AFM feedback loop. This is what causes the topographic contrast in Fig. 2c.

Figure 2d presents a zoomed-in topographic image of the patterned electrode near the "V" intersection outlined by the red-dashed box in Fig. 2c. Atomically flat terraces separated by single unit cell step heights are observed. These stepped terraces result from the STO surface miscut angle, which gets imprinted in the LAO growth, demonstrating the high quality of the epitaxially grown layer by pulsed laser deposition (PLD)[21,22]. The difference between the lighter and darker contrast regions shows the patterned area, forming a "V" shape. The patterning is much more dramatic in the CPD map in Fig. 2e. Here, the CPD value is extracted from individual frequency shift vs. sample bias, ($\Delta f$ vs. $V_{\text{Bias}}$), curves obtained at each pixel in the image of Fig. 2d. These curves display a parabolic frequency shift vs. bias where the extrema determine the CPD values (Supplementary Fig. 1). A clear gradient is observed across the patterned boundary, indicating spatial variation in the contact potential difference around the channel. A maximum CPD difference of almost 1.5 V is observed between the patterned and non-patterned areas. No CPD variation is seen between the patterned and unpatterned areas on the exposed STO surface where the LAO layer has been removed (Supplementary Fig. 3), demonstrating the critical role of the LAO layer for creating the 2DES.

**Electrostatic modeling of the carrier density profile**

KPFM measurements are sensitive to local potential variations, as schematically defined in terms of the CPD shown in Fig. 2a and b, and were used to investigate the local potential variations between the patterned (conducting) and non-patterned (non-conducting) regions of the LAO/STO heterostructure. Considering that the tip work function $\phi_{\text{tip}}$ remains constant across the LAO surface, variations in CPD directly reflect a local work function change on the LAO surface ($\phi_{\text{LAO}}$). Variations in CPD value across the sample are thus attributed to changes in the LAO potential landscape, induced by interfacial carrier accumulation and surface dipole modifications forming a capacitor with the LAO dielectric in between them[23,24].

The LAO/STO interface remains insulating in non-patterned regions, with minimal band bending, as evidenced from the conductive jump when patterned areas are joined (Fig. 1b). We then associate the change in carrier density in the patterned region, $\Delta n$, to be related to the change in work function between the patterned and non-patterned regions, as related to the change in CPD as[25,26],

$$\Delta\phi_{\text{CPD}} = \phi_{\text{LAO}}^{\text{non-patterned}} - \phi_{\text{LAO}}^{\text{patterned}} \approx \frac{e\Delta n\, d_{\text{LAO}}}{\varepsilon_0 \varepsilon_r}, \qquad (1)$$

where $e$ is the elementary charge, $d_{\text{LAO}}$ is the thickness of the LAO layer, $\varepsilon_0$ is the permittivity of free space. $\varepsilon_r$ is the dielectric constant of LAO, which we take from the literature to be $\varepsilon_r = 25$ [27,28]. We apply this relation to obtain an upper bound on the carrier density in the ULV-EBL device, as it neglects contributions from surface dipole variations.

The LAO/STO system generally exhibits a dome shape in the superconducting critical temperature vs. density phase diagram, which can be tuned by electrostatically doping the interface[29,30]. Estimates of the carrier density for superconductivity in the LAO/STO interface suggest an optimal range from $n \approx (1.5 - 6.5) \times 10^{13}$ cm$^{-2}$ [29]. We estimate the carrier density variation in the patterned region from the CPD map in Fig. 2e using equation (1) (Fig. 3a). Comparing the non-patterned and patterned areas, the estimated density changes by $\Delta n \approx 1 \times 10^{14}$ cm$^{-2}$. The line trace in Fig. 3b shows the carrier density profile across one of the patterned channels in the "Y" branch.

**Dissipation measurements**

Non-contact AFM dissipation spectroscopy detects nanoscale energy losses (dissipation) by measuring the energy loss per oscillation cycle required to maintain constant sensor oscillation amplitude. This dissipation signal reflects the dissipative response of the surfaces and interfaces to the oscillating tip's perturbations. In conducting and semiconducting systems, the dominant loss channel is electronically driven ohmic dissipation. Superconductivity modifies the losses by suppressing ordinary ohmic losses in the sample, making spatially resolved dissipation spectroscopy a local diagnostic tool for tracking and mapping superconductivity. In general, dissipation spectroscopy can reveal changes in the dominant energy-loss channel across electronic or structural transitions, as well as local electronic inhomogeneity. Despite its exquisite sensitivity to local electronic states, dissipation spectroscopy has rarely been employed to study oxide superconductors at the nanoscale[17,18].

Local dissipation measurements of Device A reveal distinct bias-dependent behaviors between superconducting and non-superconducting regions. Figure 4a shows a zoomed-in region of the bottom of the 2D channel indicated by the orange dashed box in Fig. 2c. The patterned area is observed by the higher Z contrast (brighter) in the topographic image. To investigate the origin of these signals, we perform voltage-dependent dissipation spectroscopy at two representative locations: one outside the patterned region (point B) and one inside the conducting channels (point A), as marked in Fig. 4a. The resulting spectra are plotted in Fig. 4b. These regions exhibit significant spatial variation. Higher dissipation is observed outside the patterned region, even at the sample bias corresponding to the CPD (minimum of the curve), where the electrostatic force is minimized. In contrast, significantly less energy dissipation is observed inside the patterned region.

We observe that the dissipation has a clear quadratic dependence on bias voltage ($\propto \Delta V^2$) on the dielectric background (point B), and there is a higher dissipation level both at and away from the CPD. This behavior is consistent with capacitive energy losses typically observed on metallic surfaces due to ohmic dissipation[31]. Similar ohmic losses are present on doped semiconductors[17]. The quadratic dependence arises from the scaling of ohmic loss with the magnitude of the probe-sample force, and the electrostatic force varies as $\Delta V^2$, where $\Delta V = (V_{\text{CPD}} - V_{\text{Bias}})$. In contrast, inside the patterned superconducting channels (point A), the baseline dissipation and the quadratic contribution are strongly suppressed, while a pronounced higher-order $\Delta V^4$ component emerges. As shown in Fig. 4b, it is well described by a composite fit of the form $\alpha \Delta V^2 + \beta \Delta V^4$. Together with the suppressed of $\Delta V^2$, the emergence of the higher order $\Delta V^4$ component in the patterned region indicates an additional nonlinear dissipation channel beyond simple ohmic-like linear response losses under electrostatic coupling. Similar nonlinear bias dependence has been reported in bulk superconducting Nb and attributed to phonon-mediated dissipation driven by the electronic coupling of the tip, which becomes prominent when ordinary ohmic losses are suppressed in superconducting state[31,32]. We therefore interpret the $\Delta V^4$ contribution as a robust local marker of superconductivity, consisting of a spatially resolved suppression of low energy excitations due to the quasiparticle gap opening. The presence of a smaller but nonzero $\Delta V^2$ term in the patterned superconducting regions suggests a coexistence of dissipation mechanisms. Some possible origins include residual electrostatic coupling from the finite tip size, incomplete or inhomogeneous superconductivity within the channel, and vertical tip oscillation sensitive to out-of-plane fields[23,33]. Additional contributions may also arise from quantum capacitance or nonlinear screening at the oxide interface[34].

The connection of the dissipation signals to the superconducting phase at the oxide interface is further investigated using dissipation spectroscopy maps, which are performed across a range of magnetic fields and sample biases on the 4 uc sample (Device A). The sample bias was swept from -1200 mV to +200 mV while the magnetic field was incrementally tuned from $-300$ mT to $+300$ mT. The resulting two-dimensional dissipation map reveals a strong dependence of the damping signal on both sample bias and magnetic field (Fig. 4c).

At low magnetic fields ($|B| \leq 50$ mT), the dissipation curves maintain a convex shape centered at the CPD. As shown in the line cut at -18 mT in Fig. 4d (red), the curve shows $\alpha \Delta V^2 + \beta \Delta V^4$ dependence, consistent with the zero-field point spectroscopy measurements in Fig. 4b. This shape is characteristic of superconducting regions where electrostatic energy loss is suppressed by the local environment, but not completely eliminated. As the magnetic field approaches to the critical field of ~200 mT, the dissipation changes dramatically; the curvature reverses and forms a concave feature in the dissipation curves and map (see the green curve in Fig. 4d).

The complete field-dependent evolution of the dissipation is illustrated as a magnetic field map in Fig. 4c. Figure 4e displays a vertical line cut at the CPD value, where the dissipation exhibits a minimum near zero field. The dissipation then increases with the magnetic field, reaching a maximum as it approaches the critical magnetic field, and decreases at higher fields. This evolution indicates a crossover from a low-dissipation superconducting regime to a more dissipative state, likely involving enhanced phase fluctuations due to pair breaking of the superconducting state. The enhanced dissipation around $\approx \pm 100$ mT can be attributed to the enhanced superconducting fluctuations or local phase separation[8,35].

Interestingly, the crossover (increased dissipation) occurs as the field approaches 200 mT, the critical field observed in global transport measurements presented in Fig. 1e for Device A. This demonstrates a direct connection between the dissipation signals and the superconducting state at the LAO/STO interface. The dependence of dissipation on sample bias may reflect some contribution to local doping of the system. In LAO/STO systems, superconductivity exhibits a dome-shaped dependence on carrier density, with superconductivity emerging above a threshold density, peaking at intermediate doping, and weakening at higher densities[29,30]. As a result, local deviations from optimal doping can reduce the apparent critical field and enhance phase fluctuations, which may account for the dome-shaped dissipation vs. sample bias (Fig. 4c and d).

Our results indicate that the dissipation is enhanced near the critical field. In Fig. 5, we therefore examine the dissipation across the boundary of the 2D region highlighted in the yellow dashed rectangle in Fig. 4a obtained near the critical magnetic field at -235 mT. The three-dimensional data set measures the AFM frequency shift and excitation (dissipation) as a function of bias and XY spatial position. The CPD is extracted from the frequency shift vs. bias curves and is shown in Fig. 5a. The CPD varies by 0.65 V over the rectangular area, with lower CPD and higher carrier density in the upper region corresponding to the patterned region in Fig. 4a. The dissipation results are shown in the $Y$ vs. $V_B$ plane in the middle of the rectangular region at $X = 125$ nm in Fig. 5b. The dissipation curves in the underdoped and overdoped regions display similar concave dissipation vs. bias curves (Fig. 5f red and blue curves) as observed in Fig. 4d at fields near and above the critical field of $\pm 200$ mT, implying a similar dissipation mechanism. The green curve in Fig. 5f, however, shows a convex parabolic behavior with $\Delta V^4$ dependence, resembling the green curve in Fig. 4b and the red curve in Fig. 4d.

The spatial dependence of dissipation is obtained from vertical line cuts at selected biases from the 2D map in Fig. 5b and is then expanded along the X dimension, as shown in Fig. 5c-e. A clear boundary is observed at the edge of the patterned region where the dissipated energy exhibits a minimum. Comparing the dissipation linecut (Fig. 5h) with the carrier density change (Fig. 5a) shows that the dissipation minimum region is confined to ~200 nm. This narrow channel, where energy dissipation dips to a minimum, occurs near the relative intermediate doping range (Fig. 5g), which is expected for superconductivity at the LAO/STO interface and is located near the edge of the patterned region. The green curve in Fig. 5f corresponds to the dissipation vs. bias in this channel.

Together, these results highlight how spatially resolved CPD and dissipation mapping provide complementary information: CPD measurements track electrostatics, while dissipation provides a direct local diagnostic of the superconducting response.

**Discussion**

We estimate the change in carrier density in the patterned channels by utilizing the CPD changes obtained from the KPFM. Comparing the density profile with estimates for optimal density suggests the patterned channel is slightly overdoped, and the optimal density is obtained on the sides of the patterned channel. This finding is consistent with a scenario in which 2D channels are overdoped in the center, and necessarily trace out the superconducting dome as the insulating boundary is reached[7]. The edge of an overdoped 2D channel is necessarily bounded by an optimally doped region where pairing and superconductivity are strongest. We find that the width of the superconducting channel is dictated by the electrostatic screening properties of the LAO dielectric and is independent of the patterned channel width. However, this carrier density estimate is an upper bound, as we neglected changes in the surface dipole, and the dissipation measurements provided a more distinct signature of the superconductivity in the patterned channels.

Focusing on the dissipation results, the observed $\Delta V^4$ component term at zero magnetic field provides evidence of superconductivity due to suppression of ohmic losses in the patterned channels. The optimally doped region (green curve in Fig. 5f) exhibits persistent superconducting behavior, as shown in Fig. 4b green curve. In contrast, the over-doped and underdoped regions display concave behavior, similar to Fig. 4d above the critical field (blue and purple curves). This concave behavior indicates a change in bias dependence in the overdoped and underdoped regions without global superconductivity, coexisting with spatially localized gapped superconducting regions. The concave curves in Figs. 4 and 5 indicate an enhanced dissipation at the bias voltage corresponding to the CPD. We interpret this as increased ohmic losses due to pair-breaking, driving the system toward an insulating or weakly conductive state. One possible interpretation of the concave shape of the curve away from the CPD bias is that bias values away from the CPD shift the interface doping, lowering the critical field. Since dissipation is a maximum just before the critical field values, this naturally yields a concave curve when the line cut is held constant at a magnetic field from the map in Fig. 4c.

Notably, the dissipation measurements in Fig. 4 are correlated with the critical magnetic field, consistent with macroscopic transport measurements of superconductivity in these devices. The region where the dissipation reaches a minimum (Fig. 5h) suggests a superconducting channel where ohmic losses are further suppressed. This measurement is performed at the critical magnetic field. For optimal doping, this channel likely has a higher critical field than observed in the transport measurements (Fig. 1e). We propose that the intrinsic superconductivity resides in the optimally-doped edge channels (Fig. 5h), while the superconducting signatures observed in the overdoped interior arise from a proximity effect—Cooper pairs diffusing from the edges over the superconducting coherence length ξ. This interpretation is quantitatively supported by established material parameters: the superconducting coherence length in LAO/STO has been measured to be ξ ≈ 60–100 nm [36–38]. For our ~1 μm wide channel, proximity effects extending ~ξ from each edge would influence a substantial portion of the channel interior, naturally explaining why superconducting signatures appear throughout the patterned region at zero field. This proximity effect scenario provides a compelling explanation for the spatial hierarchy observed in our field-dependent measurements: as magnetic field increases, the proximity-induced superconductivity in the interior is suppressed first (the "weakest link"), while the intrinsic edge superconductivity persists to higher fields. The interior lacks its own robust superconducting gap—it inherits superconducting correlations from the edges. When the proximity coupling is weakened by the applied field, the interior reverts to a non-superconducting state before the edges do.

**Outlook**

Our measurements are consistent with superconductivity preferentially localized near (~200 nm) channels at the edges of patterned regions. The dissipation spectra show characteristic $\Delta V^4$ nonlinear component in superconducting regions, contrasting with $\Delta V^2$ ohmic losses in normal regions. Near the critical field, the dissipation signal behavior changes strongly, consistent with the suppression of superconducting correlations and a crossover in the dominant dissipation channel. Both inside and outside the patterned channels, dissipation vs. bias curves show concave dissipation-bias dependences near and above the critical field, while the outside region shows a stronger bias dependence. Convex dependence is only preserved in the strip near the channel border, suggesting a more robust superconducting phase. These spatially resolved measurements are suggestive of the one-dimensional hypothesis, explaining why critical currents do not scale with channel width and why pairing persists beyond the bulk superconducting regime. However, more work is needed to definitively nail down this conclusion, with local measurements as a function of temperature, field, gating bias, and frequency. Importantly, this work demonstrates that this path is achievable via atomic force microscopy measurements at ultra-low temperatures, where the LAO/STO interface superconductivity resides.

**Acknowledgments:**

Certain commercial equipment, instruments, or materials are identified in this article to specify the experimental procedure adequately. Such identification is not intended to imply recommendation or endorsement by the National Institute of Standards and Technology, nor is it intended to imply that the materials or equipment identified are necessarily the best available for the purpose. D.Y. thanks Prof. E. Tosatti for the insightful discussion related to the dissipation data.

**Funding:**

University of Maryland and NIST Joint Quantum Institute Grant No. 70NANB21H126 (SK, DYildiz)

Office of Naval Research Grant No. N00014-20-1-2352 (SK, DYildiz, DYang)

Office of Naval Research Grant No. N00014-21-1-2537 (JL, CBE).

Deutsche Forschungsgemeinschaft (DFG, German Research Foundation) within Project-ID 314695032—SFB 1277 and GRK2905 (2905) (FJG)

Vannevar Bush Faculty Fellowship (ONR N00014-20-1-2844) (CBE).

Gordon and Betty Moore Foundation's EPiQS Initiative, Grant GBMF9065 (CBE).

**Author contributions:**

JL and JAS conceived the idea and designed the scope of the experiment. JL and JAS supervised the project. DYildiz, SK, MY, DYang carried out the measurements and analyzed the data. KL, RS, CBE, FJG, DYang, MY, EMS, PI contributed to fabricating the devices for the project. DYildiz, DYang, JAS wrote the original draft of the manuscript, and all authors discussed and provided final editing of the manuscript.

**Competing interests:** FJG holds patents on the qPlus sensor. All other authors declare that they have no competing interests.

**Data availability:** All data are available in the main text or the supplementary materials. Further information is available from the corresponding authors on request.

**Additional Information**

**Supplementary Information file**

**Correspondence and requests for materials** should be addressed to JAS or JL.

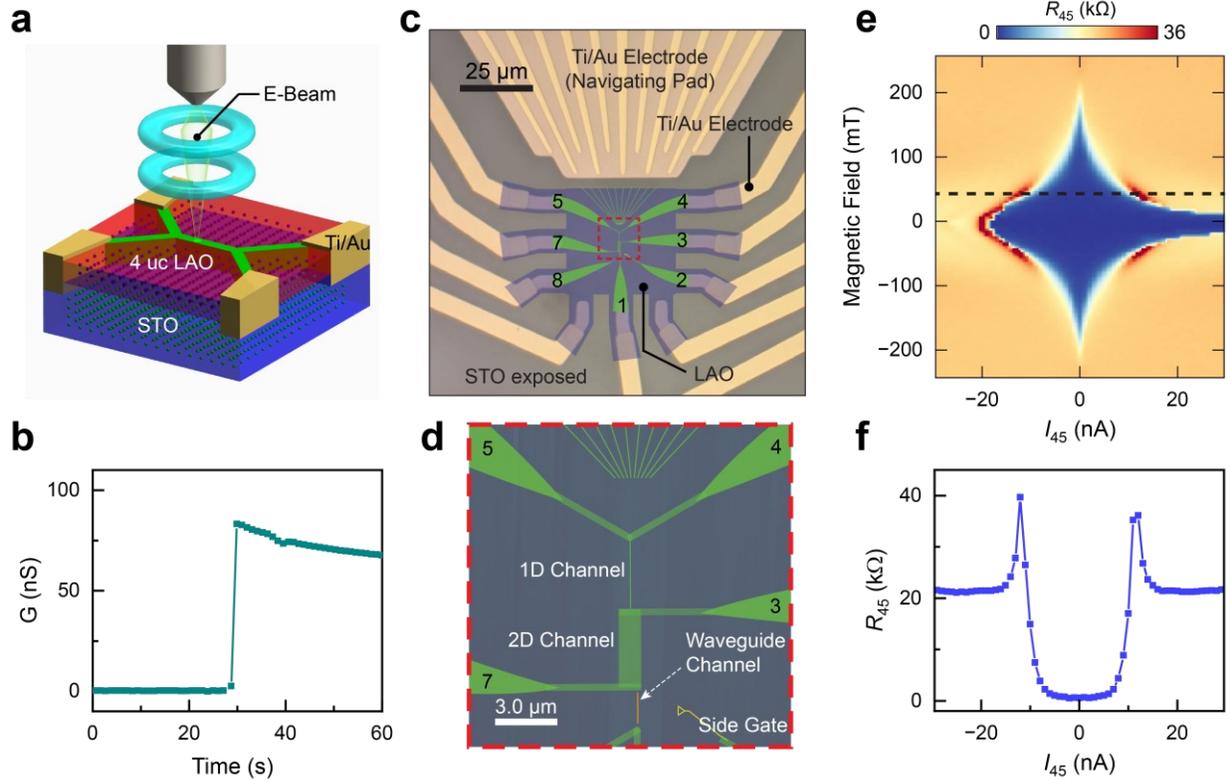

**Fig. 1. Patterning the LAO/STO interface with superconducting channels**. (**a**) Schematic diagram of ULV-EBL patterning. (**b**) Conductance measurement evolution during patterning. A sharp conductance increase is observed when the pattern connects two contacts. (**c**) Optical microscope image of the device. The top LAO layer has been etched away outside the dark blue region. The lithographically-defined area patterned by ULV-EBL is indicated in green. (**d**) The region enclosed by the dashed red box in (c) is magnified to show the various patterns. The dimensions of the patterned areas are: 1D channel (l × w): 3.5 µm × 10-20 nm (single line), 2D channel (l × w): 3.5 µm × 1 µm, and the waveguide: 1.35 µm channel with 100 nm gap separation on each side. (**e**) Magnetic field dependence of channel resistance measured through contacts 4 and 5 marked in (d), as illustrated in (Supplementary Fig. 5). The asymmetric dependence on current likely reflects asymmetric contact transparency or a built-in electrostatic asymmetry, neither of which affects the superconducting properties probed here. (**f**) Line cut of the channel resistance at 47 mT indicated by the dashed line in (e).

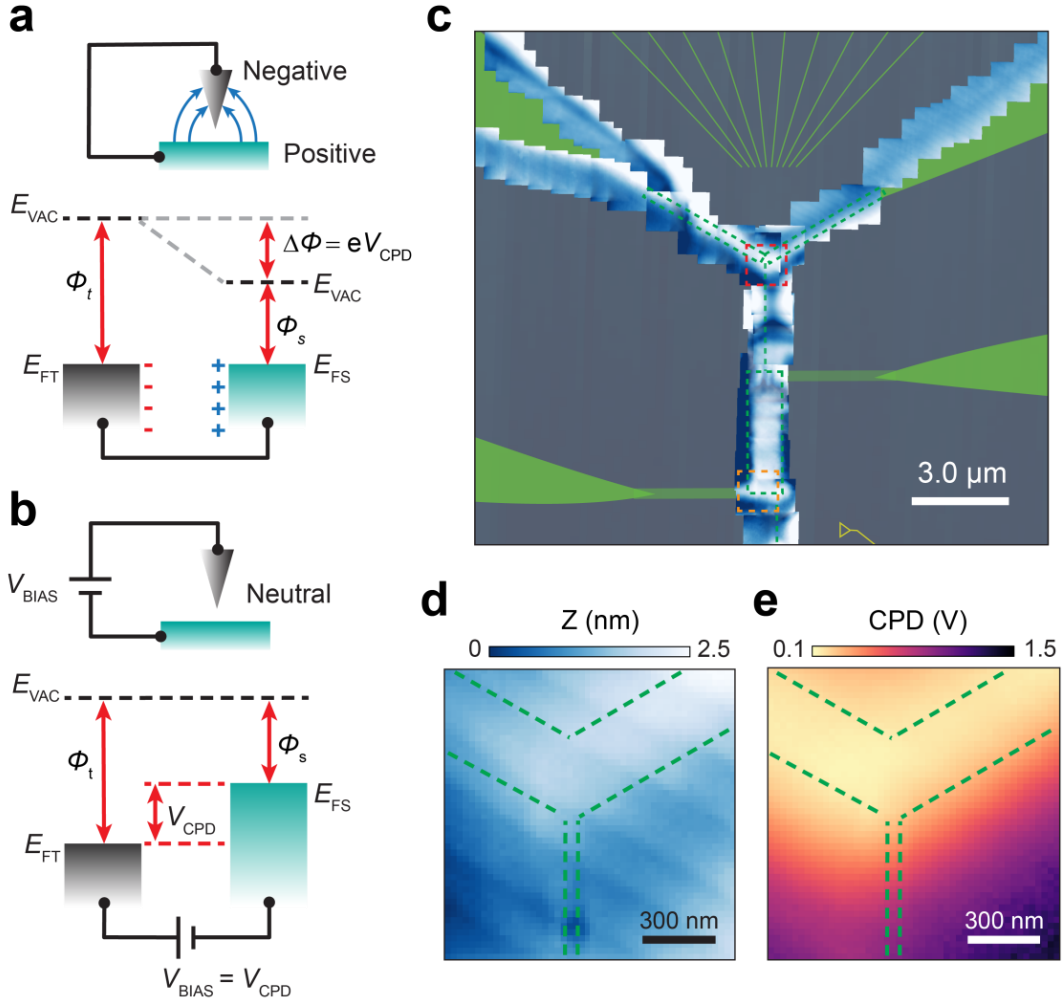

**Fig. 2. Visualizing the patterned channel**. Schematic diagrams illustrating the concept of contact potential difference (CPD). (**a**) When the tip and the sample are electrically in contact, a potential difference across the vacuum gap arises due to the difference in their work functions. (**b**) A bias voltage $V_{Bias}$ is applied to the sample to nullify this difference, making the tip and the sample electrostatically neutral and $V_{Bias}$ is equivalent to the contact potential difference. (**c**) AFM topography along the patterned region of the four uc LAO/STO interface, acquired in frequency modulation (FM) mode with an oscillation amplitude of 3.2 nm –5.0 nm. (**d**) AFM topography and (**e**) contact potential difference map within the red-dashed boundary indicated in (c), respectively. Diagonal stripes in (d) are atomic terraces of the LAO. CPD is extracted from $\Delta f$ vs. $V_{Bias}$ parabolic curves at each grid point (see Supplemental Information). The patterned region boundaries are indicated in green dashed lines in (c-e). A 4.0 nm oscillation amplitude with frequency shift $\Delta f = -3.0$ Hz was used for feedback of the measurement in (d,e). $T$ = 10 mK.

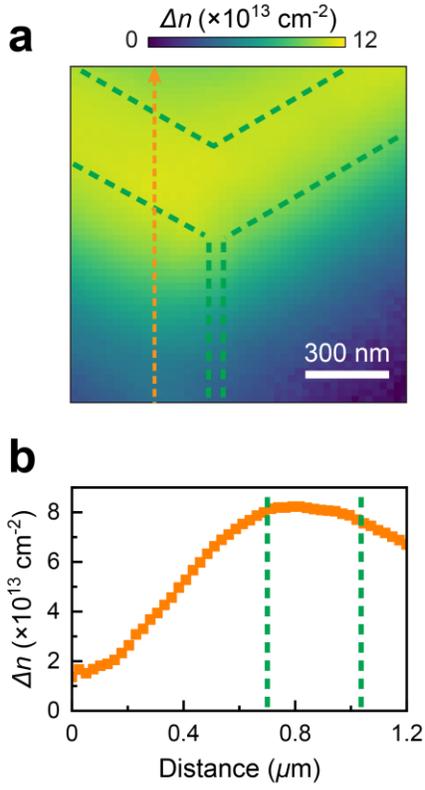

**Fig. 3. Estimation of spatial carrier density profile in a superconducting channel**. (**a**) Carrier density map of the patterned LAO/STO region calculated from the CPD map in Fig. 2e using equation (1). A CPD of 1.024 V was used for the non-patterned area obtained at the bottom left corner of the image in Fig. 2e. (**b**) Charge density profile across the pattern boundary extracted along the orange dotted line in (a). The statistical uncertainty is less than the symbol size. The estimated pattern boundaries are indicated by green dashed lines.

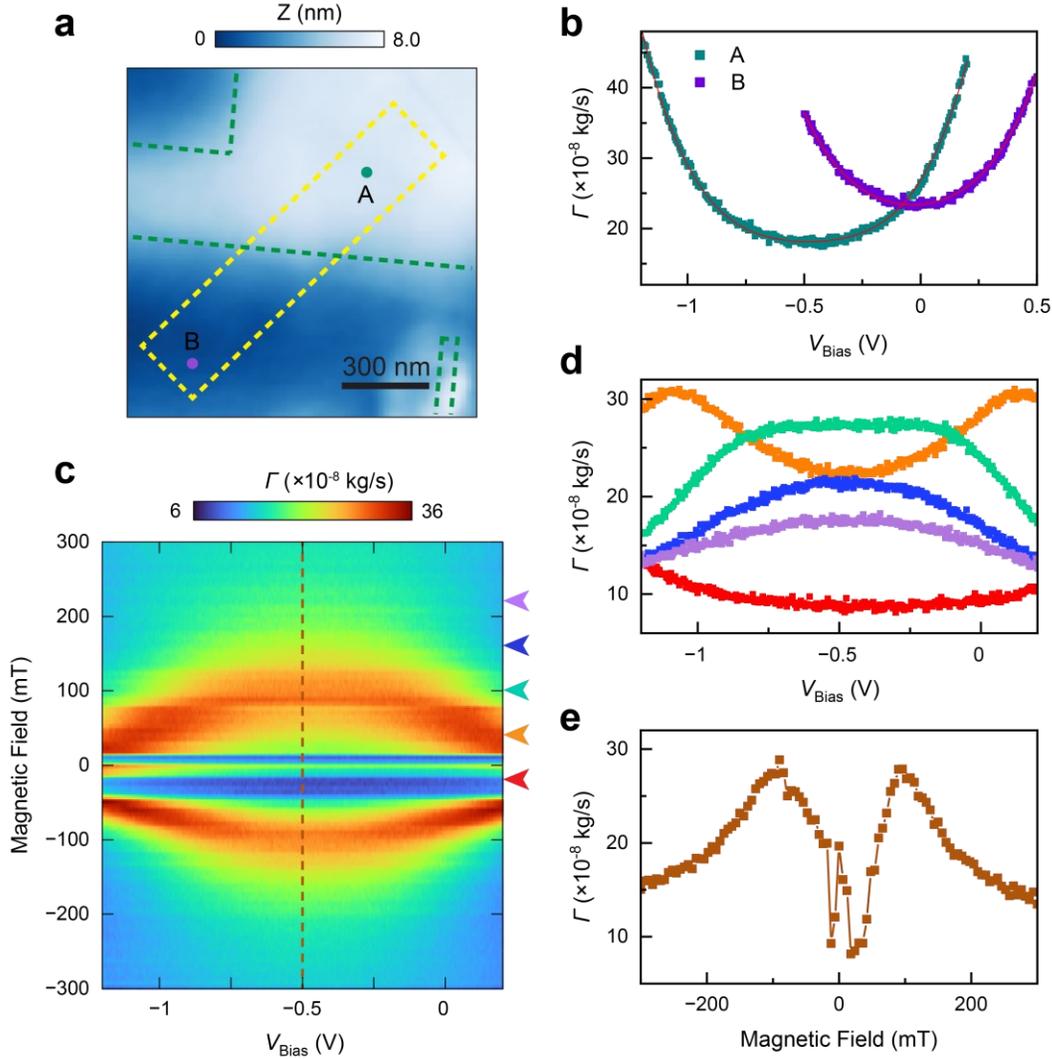

**Fig. 4. AFM energy dissipation measurement at LAO/STO interface**. (**a**) AFM Topography image at the end of the 2D channel region of the 4 uc LAO/STO device (see orange box in Fig. 2c). The green dashed lines indicate the estimated patterned boundary. (**b**) Damping coefficient vs. sample bias curves obtained at the two points marked in (a). The two curves are fit to the form $\alpha \Delta V^2 + \beta \Delta V^4$. The ratios of the 4$^{\text{th}}$ to 2$^{\text{nd}}$ order coefficients, $\beta/\alpha = (3.85 \pm 0.03) V^{-2}$ and $(1.84 \pm 0.01) V^{-2}$ for curves A and B, respectively. The uncertainty estimates represent one standard deviation derived from the non-linear least squares fits. The measurements in the patterned area have the larger 4th-order components. (**c**) AFM dissipation vs. magnetic field and sample bias obtained near point A in (a). (**d**) Horizontal line cuts from (c) at magnetic fields of -18 mT, 42 mT, 102 mT, 162 mT, and 222 mT (indicated by colored arrowheads). (**e**) Vertical line cut at the sample bias of -500 mV showing the dissipation field dependence at the CPD. A glitch occurred in the measurement at a small negative field due to the AFM feedback. Otherwise, this measurement is symmetric in the magnetic field. A 3.2 nm oscillation with frequency shift $\Delta f = -1.0$ Hz was used for feedback. The statistical uncertainty for (b), (d), and (e) is less than the symbol size. $T$ = 10 mK.

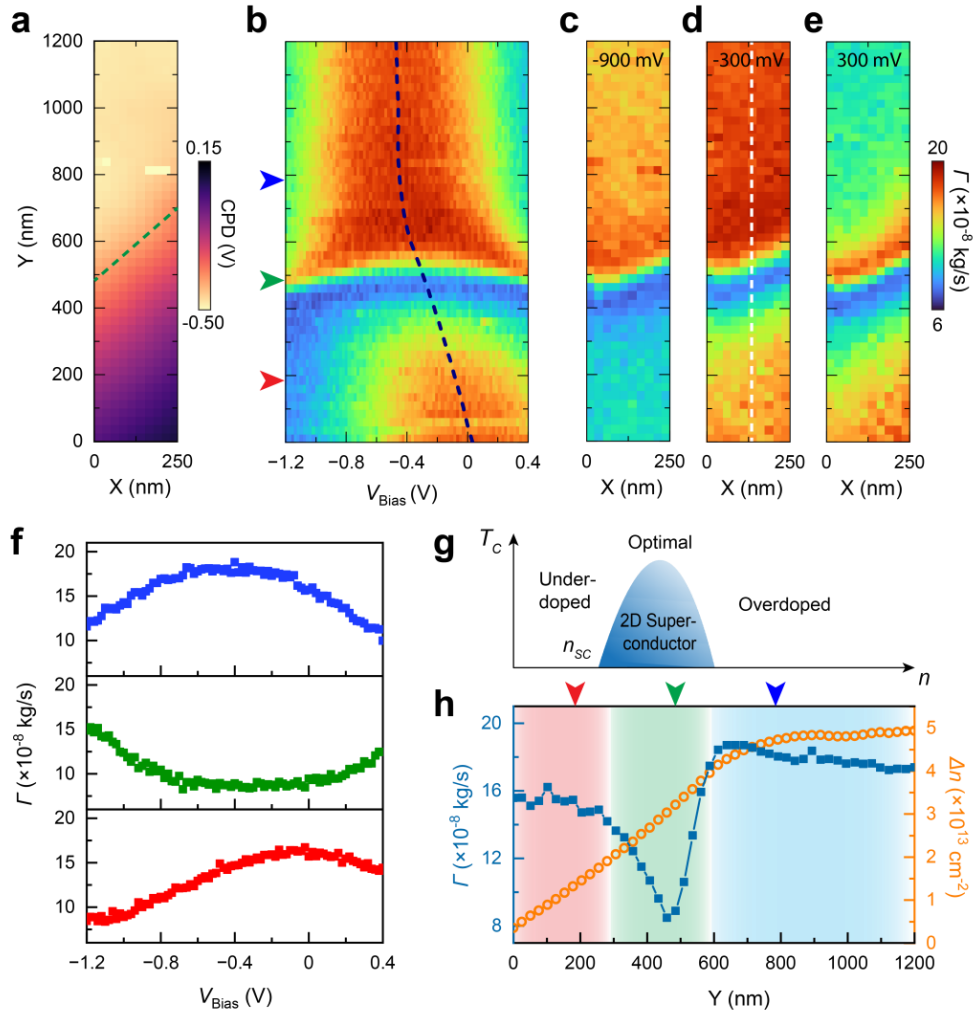

**Fig. 5. Spatial mapping of energy dissipation across a patterned channel boundary**. (**a**) CPD map extracted from $\Delta f$ vs. $V_{Bias}$ parabolic curves at each grid point in the rectangular region outlined by the dashed yellow line in Fig. 4a. The green dashed lines indicate the estimated patterned boundary. (**b**) The damping coefficient in the $Y$ vs. $V_{Bias}$ plane obtained at $X = 125$ nm, obtained from the AFM excitation vs. sample bias data. The dark blue dashed line shows the evolution of the CPD as the spatial map crosses the patterned boundary. (**c-e**) Spatial maps in the $XY$ plane of the damping coefficient at selected sample biases of -900 mV, -300 mV, and 300 mV, respectively. Note the narrow strip where the damping coefficient dips to minimum values between $Y$ values of 400 nm to 600 nm. (**f**) Line traces of the damping coefficient vs. sample bias at the selected $Y$ values indicated by the colored arrows in (b). The statistical uncertainty is less than the symbol size. (**g**) Schematic of the superconducting phase diagram of the superconducting critical temperature $T_c$ vs density $n$ for the LAO/STO system. (**h**) A comparison between the damping coefficient obtained by the dashed line in (d) and the carrier density estimated from the CPD map in (a) at $X = 125$ nm using equation (1). A CPD of 0.07 V was used for the non-patterned area obtained from the bottom right corner of the CPD map in (a). A 3.2 nm oscillation with frequency shift $\Delta f = -0.800$ Hz was used for feedback. The statistical uncertainty for (f) and (g) is less than the symbol size. $B$ = -235 mT. $T$ = 10 mK.

# Supplementary Information for

## Local probing of superconductivity at oxide interfaces with atomic force microscopy

Dilek Yildiz[1,2,3,*,†], Sungmin Kim[1,2,*], Dengyu Yang[1,2,4,*] Muqing Yu[4,*], Kyoungjun Lee[5], Ruiqi Sun[5], En-Min Shih[1,6], Steven R. Blankenship[1], Patrick Irvin[4], Franz J. Giessibl[7], Chang-Beom Eom[5], Jeremy Levy[4], and Joseph A. Stroscio[1,†]

[1]Physical Measurement Laboratory, National Institute of Standards and Technology; Gaithersburg, MD 20899, USA.

[2]Joint Quantum Institute, Department of Physics, University of Maryland; College Park, MD 20742, USA

[3]Department of Advanced Material Science, The University of Tokyo; Chiba 277-8561, Japan.

[4]Department of Physics and Astronomy, University of Pittsburgh; Pittsburgh, Pennsylvania 15260, USA

[5]Department of Materials Science and Engineering, University of Wisconsin–Madison; Madison, Wisconsin 53706, USA

[6]Department of Chemistry and Biochemistry, University of Maryland; College Park, MD 20742, USA.

[7]Institute of Experimental and Applied Physics, University of Regensburg; Regensburg 93040, Germany

[*]These authors contributed equally to this work.

[†]Corresponding authors: yildiz@g.ecc.u-tokyo.ac.jp, joseph.stroscio@nist.gov

**The PDF file includes:**

Materials and Methods

Supplementary Text

Supplementary Figs. 1 to 5

References (*39-44*)

## Methods

### Device fabrication methods

A commercially available SrTiO$_3$ (STO) substrate is treated with buffered hydrofluoric acid (BHF) etching to produce an atomically flat TiO$_2$-terminated surface. A thin LaAlO$_3$ (LAO) film is then grown on top by pulsed laser deposition (PLD), with the film thickness precisely monitored by in situ reflection high-energy electron diffraction (RHEED) oscillations[2,22]. Two devices were investigated: Device A with 4 unit cells (uc) of LAO (ULV-EBL patterned) and Device B with 6 uc of LAO (c-AFM patterned).

The devices were patterned at the University of Pittsburgh and then subsequently placed in a vacuum carrier. The vacuum carrier and devices were transported to NIST within several hours. After arriving at NIST, the devices were briefly exposed to the atmosphere to load them into the vacuum load lock of the mK STM system. At all times, the devices were kept from ambient lighting, and only red-filtered lighting was used to view the devices as they were transported through the series of vacuum systems that comprise the mK STM system[39]. The devices for measurements were then placed in a custom scanning probe module capable of simultaneous AFM, STM, and magnetotransport measurements[40]. The system utilizes a dilution refrigerator which operates at a base temperature of 10 mK with magnetic fields up to 15 T perpendicular to the sample plane. Multimodal measurements are accomplished by using custom-designed sample and probe tip holders, which feature eight electrical contacts for devices and probe sensors, allowing for multimodal measurements of STM, AFM, and magnetotransport measurements in a single system operating at temperatures down to 10 mK[40]. The qPlus AFM sensor uses a design that incorporates an integrated excitation electrode on the sensor. For the qPlus sensor, two contacts were wired to read out the AFM sensor, one for the STM tunneling current, and one for the sensor excitation. The probe tips were fabricated from 100 μm diameter Ir wire.

### C-AFM patterning

C-AFM lithography[41] was performed using a commercial AFM system. The patterning process was conducted at room temperature in a controlled environment with relative humidity maintained between 40–50%. An Aspire conical force modulation tip with doped single-crystal silicon was used for both imaging and lithography. AFM imaging was carried out in AC (tapping) mode at a resonant frequency of approximately 75 kHz. Nanoscale patterning was performed in contact mode with a tip bias of +10 V. A custom electrical feedthrough setup was integrated to allow for in situ transport measurements during the patterning process.

### ULV-EBL patterning

ULV-EBL[5,6] was performed using a commercial scanning electron microscope (SEM) system equipped with a double condenser column. The column incorporates two condenser lenses, enabling high-resolution imaging and patterning at low electron beam energies. Lithography was carried out using a high-frequency beam blanker in combination with a pattern generator hardware and control software customized to the SEM. A homemade electrical feedthrough was implemented to enable in situ transport measurements during electron beam exposure.

**Scanning Probe Microscopy (SPM) Methods**

Frequency-modulation non-contact AFM

Topography and electrostatic measurements were performed using frequency-modulation non-contact atomic force microscopy (FM-NC-AFM) and Kelvin probe force microscopy (KPFM). All measurements were conducted using an ultra-low-temperature SPM system operating under ultra-high vacuum (UHV) conditions and at cryogenic temperatures. Chemically etched Ir probe tips mounted on qPlus sensors were used for all scanning probe measurements. The sensor used for measurements had a resonance frequency of ≈18 kHz and a Q-factor of ≈100 k.

In FM-NC-AFM, topography is recorded by maintaining a constant frequency shift ($\Delta f$) while adjusting the tip-sample distance accordingly[42,43]. The apparent height in the resulting maps reflects not only the physical surface morphology but also variations in the local tip-sample force gradient, which includes contributions from electrostatic forces[23,42]. These electrostatic interactions arise from differences in the local dielectric environment and work function variations across the sample[23].

To map local electrostatic potentials, FM-KPFM was employed to measure the CPD between the tip and the sample. Local CPD is defined as: $CPD = \phi_{\text{tip}} - \phi_{\text{sample}}$, where $\phi_{\text{tip}}$ and $\phi_{\text{sample}}$ are the work functions of the tip and the sample, respectively. This allows for spatially resolved mapping of the surface potential, which is sensitive to local electronic structure and carrier distribution[23].

The electrostatic force gradient influencing the frequency shift is given by

$$F_z^{\text{elec}} \propto \frac{1}{2} \frac{dC}{dz} (V_{\text{CPD}} - V_{\text{Bias}})^2,$$

where $C$ is the tip-sample capacitance, $\frac{dC}{dz}$ is the derivative of capacitance with respect to tip-sample distance, and $V_{\text{CPD}} - V_{\text{Bias}}$ is the voltage difference between the tip and the sample. Lateral variations in this force gradient modulate the tip's vertical position to maintain a constant frequency shift, $\Delta f$, resulting in apparent height differences that are influenced by electrostatic contrast rather than actual surface topography.

The sensitivity of FM-NC-AFM to electrostatic forces depends strongly on the local dielectric properties of the sample. In regions with thinner dielectric layers, such as thinner LAO, the larger *dC/dz* enhances electrostatic interactions and makes CPD contrast visible in topography scans. Conversely, in regions with thicker dielectrics, reduced *dC/dz* weakens this contribution, limiting electrostatic influence on topographic contrast.

The CPD values are extracted from fitting the parabolic dependence of the frequency shift vs. bias curves, as illustrated in Supplementary Fig. 1 for the CPD map in Fig. 2e of the main text.

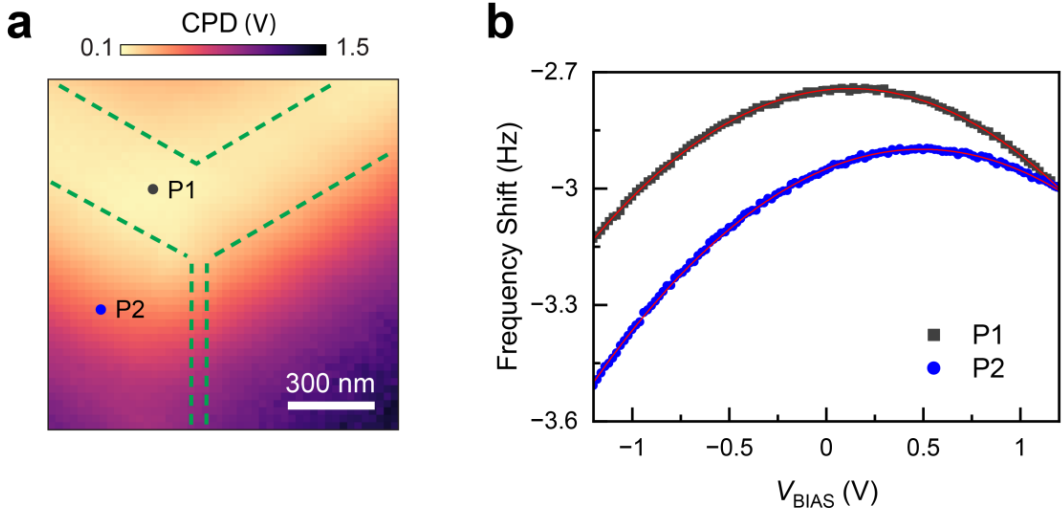

**Supplementary Fig. 1. CPD determination from frequency shift vs. $V_{Bias}$ curves**. (a) CPD map from Fig. 2e of the main text. The CPD values were obtained from fitting each $\Delta f$ vs. $V_{Bias}$ in the three-dimensional AFM data set, as illustrated in (b). (**b**) Two $\Delta f$ vs. $V_{Bias}$ curves at points 1 and 2 are fit to parabolic functions to determine their extrema values, which are the CPD.

Dissipation

    Local non-contact dissipation spectroscopy was performed using a qPlus sensor to investigate dissipated energy associated with superconducting channels. All measurements were carried out in frequency-modulation (FM) NC-AFM mode. A metallic tip mounted on a the sensor was oscillated perpendicular to the sample surface, and the excitation signal required to maintain a constant oscillation amplitude was recorded as a measure of tip-sample energy dissipation. In the absence of tip-sample interactions, the sensor oscillates in free space, and the excitation voltage needed to sustain its motion is determined solely by the mechanical properties of the sensor. This baseline behavior is independent of the applied tip bias. The intrinsic damping coefficient of the sensor in free space is given by,

$$\Gamma_0 = \frac{k}{2\pi f_0 Q}$$

where $k$ is the spring constant, $f_0 = \omega/2\pi$ is the resonance frequency and Q is the quality factor of the sensor.

    When the tip approaches the surface, additional energy dissipation arises due to tip-sample interactions such as viscoelastic damping or electronic friction. This leads to an increase in the excitation voltage $V_{exc}(z)$, required to maintain a constant oscillation amplitude. The corresponding damping coefficient under the interaction is calculated using

$$\Gamma = \Gamma_0 \left(\frac{V_{exc}(z)}{V_{exc,0}} - 1\right)$$

where the $V_{exc,0}$ is the excitation voltage in free space.

    Dissipation spectra were collected at various tip biases and magnetic fields to study the dependence of local energy loss mechanisms on electrostatic environment and external field conditions.

## c-AFM-patterned Device B

The second Device B was fabricated using the c-AFM method, as shown in Supplementary Fig. 2. During the pattern evolution, a jump in conductance occurs when the channels are formed.

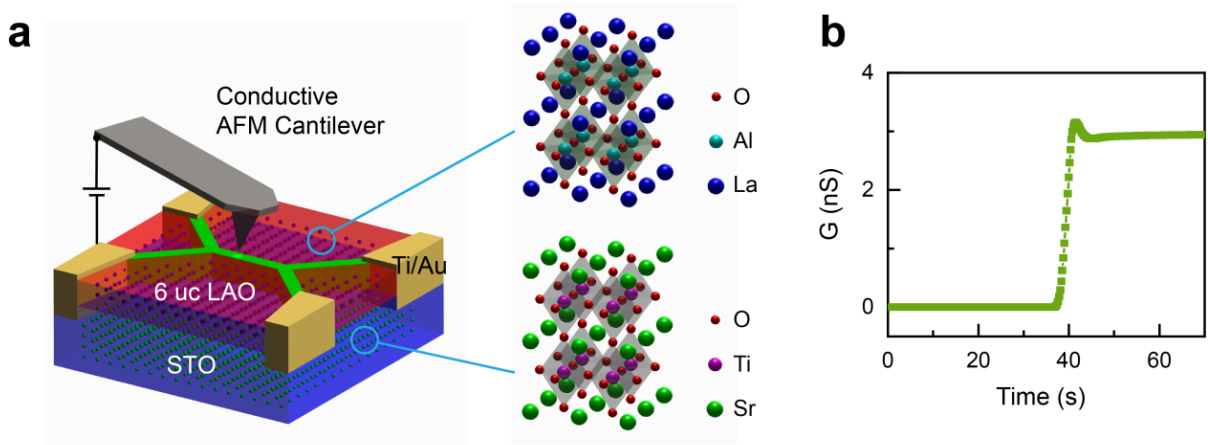

**Supplementary Fig. 2. c-AFM patterning of superconducting channels at the LAO/STO interface**. (**a**) Schematic diagram of ULV-EBL patterning and perovskite crystal structure of STO and LAO. $MO_6$ ($M$=Al, Ti) octahedra are highlighted in gray. (**b**) Conductance measurement evolution during patterning. A sharp conductance increase is observed when the pattern connects two contacts.

Selected regions of the LAO layer are etched using Ar+ ion milling, and Ti/Au electrodes are deposited to form electrical contacts to the 2DES at the LAO/STO interface. Additionally, a square region with the LAO layer was removed by inductively coupled plasma reactive ion etching (ICP-RIE) to prepare a reference area without a 2DES (sky blue areas in Supplementary Fig. 2a)[4].

Supplementary Fig. 3b shows the topography of a representative region that was measured on the c-AFM patterned 6uc LAO/STO sample (Device B) using frequency-modulated non-contact AFM. This region includes four distinct areas: (1) LAO/STO with patterning, (2) LAO/STO without patterning, (3) bare STO with patterning, and (4) bare STO without patterning. These areas are clearly distinguishable in the AFM topography image shown in Supplementary Fig. 3b. To distinguish these areas more quantitatively, we obtained the measurement of the surface work function by measuring the CPD. The CPD is obtained in the KPFM mode, where the sample bias is adjusted to minimize the force between the sample and probe. The acquired KPFM CPD map across various sub-regions of Supplementary Fig. 3b, labeled Section 1 and Section 2, is shown in Supplementary Fig. 3c. A clear contrast is observed between the patterned and unpatterned regions in the presence of the LAO layer [Supplementary Fig. 3c (Section 1)] with a CPD difference of ≈0.3 V. A decrease in CPD corresponds to an increase in work function, implying an increase in carrier density at the LAO interface. In contrast, no significant CPD variation is detected between patterned and unpatterned regions on bare STO [Supplementary Fig. 3c (Section 2)]. This suggests that the electronic response to patterning occurs only when the LAO layer is present. Without patterning, the LAO layer is below the critical thickness required to induce a 2DEG under unmodified conditions. As such, no 2DEG forms until surface protonation—introduced by patterning—shifts the conduction band minimum of STO at the interface below the Fermi level, enabling the formation of the 2DEG. In the absence of the LAO overlayer, no 2DEG is observed, regardless of whether patterning is applied [Supplementary Fig. 3c (Section 2)]. The presence of surface charge following patterning along with the 2DEG is essential for generating measurable contrast in the CPD measurements, as it gives rise to an increase in work function, consistent with the schematic shown in Fig. 2a and b of the main text.

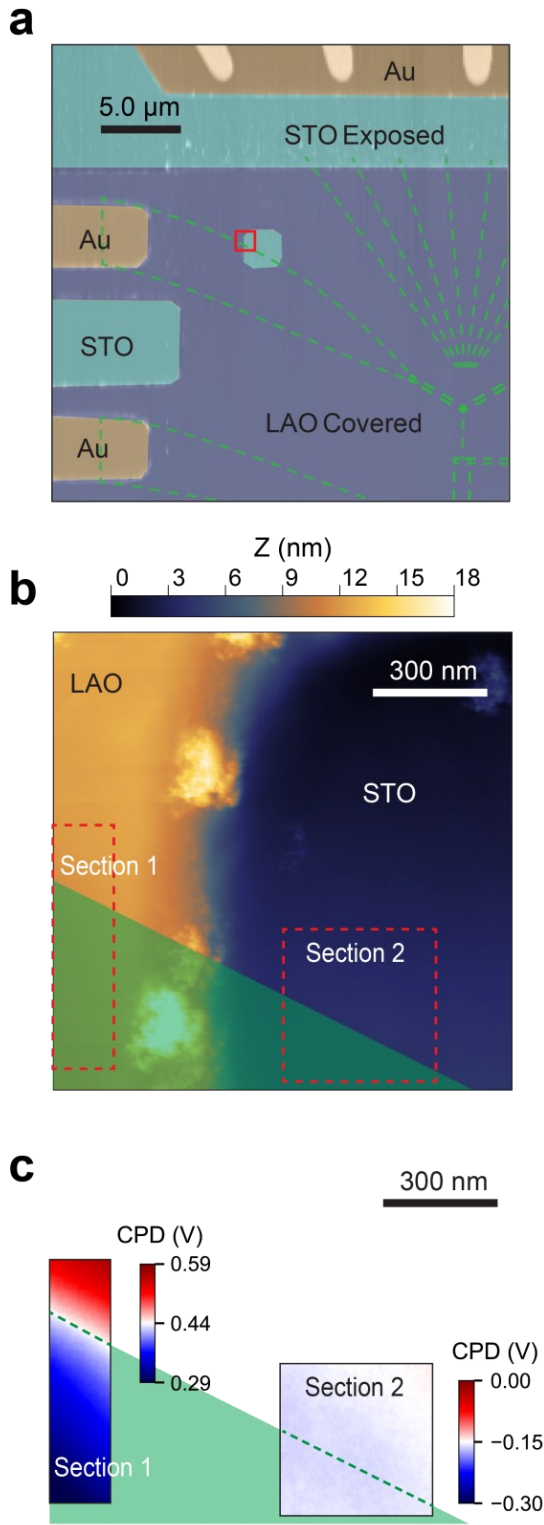

**Supplementary Fig. 3 AFM measurement of Device B patterned regions**. (**a**) AFM image of the 6 uc LAO/STO device. False color is overlaid to aid distinction. The dark blue area represents the 6 uc LAO film, while the light blue area corresponds to the etched LAO region with exposed STO. (**b**) AFM topography of an LAO-etched region as indicated by the red rectangular spot in (a). The c-AFM patterned area is marked by a green triangle. Topography was measured with a 4 nm oscillation amplitude and a -350 mHz $\Delta f$ setpoint. (**c**) CPD map of the regions shown in (b). CPD values were extracted from $\Delta f$ vs. $V_{Bias}$ parabolic curves at each point, measured with a 4 nm oscillation amplitude and a −500 mHz $\Delta f$ setpoint (see Supplementary Fig. 1). A clear contrast is observed between the patterned and unpatterned regions in the presence of the LAO layer (Section 1) with a CPD difference of ≈0.3 V. When the LAO is removed, no CPD variation is observed (Section 2), indicating there is no 2DEG at the STO surface.

Device B electrical characterization

Device B was characterized with electrical measurements to determine critical current and field as well as transconductance, as shown in Supplementary Fig. 4. From Supplementary Fig. 4a and b, the critical current was ≈ 30 nA and critical field ≈ 55 mT. A four-terminal transconductance map is recorded by varying the source-drain bias and the side gate voltage, revealing the finite-bias spectroscopy measurements (Supplementary Fig. 4c) under zero to finite magnetic field. Diamond-shaped features are observed across the full range of side gate voltages, consistent with ballistic transport in the nanowire waveguide that populates the subbands. Loop-like structures are also visible at zero field, indicative of Andreev bound states[44]. When a finite magnetic field is applied, the diamond structure splits, indicating the breaking of electron Cooper pairs compared to the zero-field measurements. These results are consistent with similar nanostructures fabricated in previous measurements, indicative of the quality of the patterned 2DES[19,20].

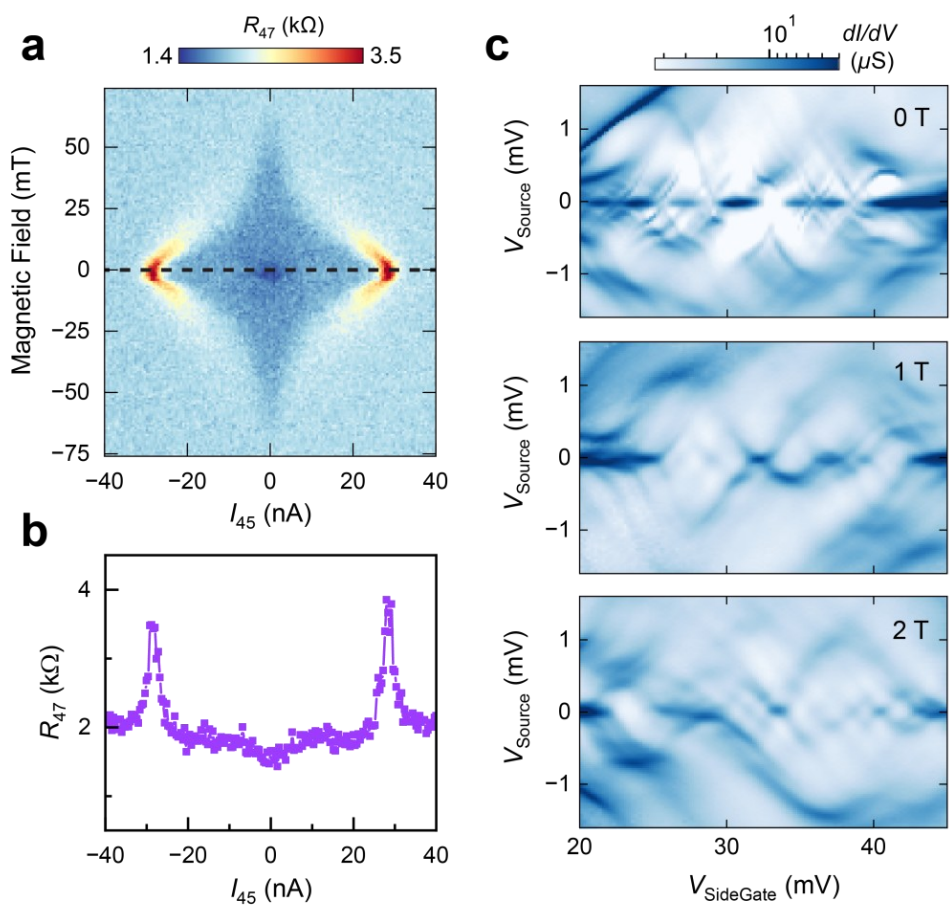

**Supplementary Fig. 4. Electronic transport measurements of superconducting channels at the LAO/STO interface in Device B**. (**a**) Magnetic field dependence of the channel resistance in a c-AFM patterned six uc interface. (**b**) Line cut of the channel resistance at 0 mT. (**c**) Electronic transport through the waveguide channel of the c-AFM patterned six μc LAO/STO interface at selected magnetic fields. Conductance between contacts 8 and 7 is measured while a source current is applied between contacts 1 and 5 (see Supplementary Fig. 5).

Electrical Schematics

    The transport measurements were obtained by creating a pseudo-current source using a voltage source and a 1 MΩ resistor. The electrical schematics for Devices A and B are shown in Supplementary Fig. 5a and 5 b, respectively.

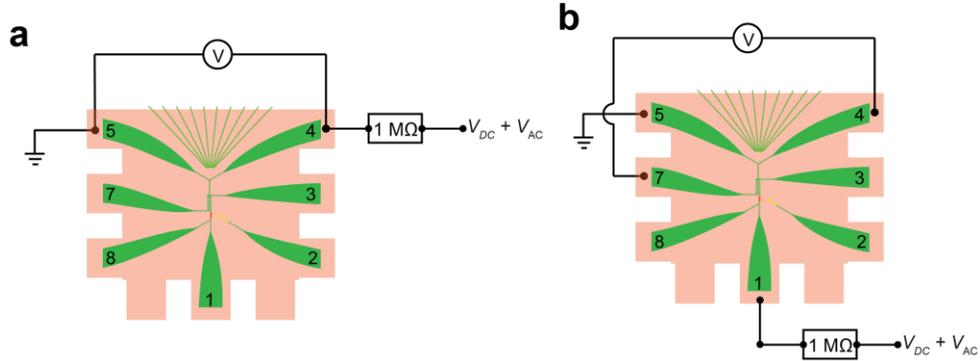

**Supplementary Fig. 5. Electronic schematics of measurements of the superconducting channels at the LAO/STO interface in Device A and B**. (**a**) Electrical schematic for the measurements shown in Fig. 1e and f of the main text. The AC voltage source was modulated at a frequency of 51 Hz and an RMS amplitude of 200 µV. (**b**) Electrical schematic for the measurements shown in Supplementary Fig. 4a and b. The AC voltage source was modulated at a frequency of 13 Hz and an RMS amplitude of 200 µV.